\documentclass[12]{article}

\renewcommand{\title}[1]{\begin{center}\bf\Large #1\end{center}}

\renewcommand{\author}[1]{\begin{center}\large #1\end{center}}

\usepackage{latexsym,amsfonts}

\begin{document}

\title{Joint Description of Periodic SL(2,R) WZNW Model and Its
  Coset Theories}

\author{
 George Jorjadze${}^a$ 
\footnote{email: \tt jorj@rmi.acnet.ge}
 and Gerhard Weigt${}^b$  
\footnote{Talk presented at the XXIXth International 
Colloquium on Group Theoretical Methods in Physics, July 15-20, 2002, 

\hspace{1.8mm}
Paris, France.}
\footnote{email: \tt weigt@ifh.de} \\
{\small${}^a$Razmadze Mathematical Institute,}\\
  {\small M.Aleksidze 1, 380093, Tbilisi, Georgia}\\
{\small${}^b$DESY Zeuthen, Platanenallee 6,}\\
{\small D-15738 Zeuthen, Germany}}

\begin{abstract}

\noindent
Liouville, $SL(2,R)/U(1)$ and $SL(2,R)/R_+$
coset structures are completely described by gauge invariant
Hamiltonian reduction of the  $SL(2,R)$ WZNW theory.
\end{abstract}

\section{Introduction}

Wess-Zumino-Novikov-Witten (WZNW) models are fascinating
two-dimensional conformal field theories with reach symmetry and
dynamical structures. Their cosets form important classes of
integrable theories. An outstanding example is the $SL(2,R)$ WZNW
model which is related to the Liouville theory and other cosets
with interesting ('black hole') space-time properties. Using
Hamiltonian reduction one can show what happens under this reduction
with the $SL(2,R)$ WZNW fields, the symplectic structure, the
Kac-Moody currents, the Sugawara energy-momentum tensor, and most
importantly how the general $SL(2,R)$ solution reduces to those of the
gauged theories. Free-field parametrisation allows canonical
quantisation, and we present as a typical quantum result a causal
quantum group commutator.

The talk is based on a series of papers \cite{OW}-\cite{FJW2}.

\section{SL(2,R) WZNW theory}

WZNW models are invariant under left (chiral) and right (anti-chiral)
multiplications of the WZNW field $g(z,\bar z)$. These Kac-Moody
symmetries provide integrability of the theory and give its general
solution as a product of chiral and anti-chiral fields, $g(z)$
respectively $\bar g(\bar z)$, which for periodic boundary conditions
in $\sigma$ have monodromies $g(z+2\pi)=g(z)M$ and $\bar g(\bar z
-2\pi)=M^{-1}\bar g(\bar z)$ with $M\in SL(2,R)$. $z=\tau +\sigma$,
$\bar z=\tau -\sigma$ are light cone coordinates.  The conformal
symmetry is generated by the traceless Sugawara energy-momentum tensor.

The basic (anti-)chiral Poisson brackets follow by inverting the
corresponding symplectic form of the $SL(2,R)$ theory. But piecing
together the chiral and anti-chiral results surprisingly simple causal
non-equal time Poisson brackets follow for the $SL(2,R)$ WZNW fields 
\cite{FJW2}
\begin{equation}\label{PB-g-g}
\{\,g_{ab}(z,\bar z),\, g_{cd}(y,\bar y)\,\}=\frac{\gamma^2}{4}\,
\,\Theta\,[2 g_{ad}(z,\bar y)\,g_{cb}(y,\bar z)-
g_{ab}(z,\bar z)\,g_{cd}(y,\bar y)],
\end{equation}
where $\Theta =\epsilon(z-y)+\epsilon(\bar z-\bar y)$ is the causal
factor.  The stair-step function $\epsilon (z)= 2n+1~$ for $~2\pi n
<\,z\,<\, 2\pi (n+1)$ ensures that causality.  This Poisson bracket
was derived for hyperbolic monodromy in the `fundamental' interval
$z-y\in(-2\pi, 2\pi)$ \, where $\epsilon(z-y)=sign\,(z-y)$, and it
holds therefore also on the line. Eq.(\ref{PB-g-g}) can be
generalised even outside this domain \cite{JW}.

From (\ref{PB-g-g}) one can derive, e.g., the canonical Poisson
brackets, the Kac-Moody algebra and
Poisson bracket relations of the energy-momentum
tensor with itself or any other field. 

The causal Poisson brackets encode the full WZNW dynamics.

\section{Coset theories}

Gauging the SL(2,R) WZNW theory with respect to the three different
types of one-dimensional subgroups $~h=e^{\alpha t}\in SL(2,R)$
\begin{equation}\label{H}
 e^{\alpha t_0}=\left( \begin{array}{cr}
  \cos\alpha&-\sin\alpha\\\sin\alpha&\cos\alpha \end{array}\right),
~~~e^{\alpha t_2}  =\left( \begin{array}{cr}
  e^\alpha&~0~~\\0&~e^{-\alpha} \end{array}\right),~~~
e^{\alpha t_+} =\left( \begin{array}{cr}
  1&0\\\alpha&1 \end{array}\right),
\end{equation}
and considering axial respectively vector gauging $~g\mapsto hgh~$ and $~
g\mapsto hgh^{-1}$ one finds six integrable \cite{MW} coset theories.
The subgroups are called compact ($t=t_0$), non-compact ($t=t_2$) and nilpotent
($t=t_+$), where the $t_n$ are  elements of the $sl(2,R)$ algebra given
by the Pauli matrices $~t_0=-i\sigma_2$, $~t_1=\sigma_1$,
$~t_2=\sigma_3$, and $~t_+=t_0+t_1$ with $~t_+^2=0$ is a nilpotent element.
The cosets will be described in terms of gauge invariant components
of the WZNW field. Taking into account the condition $~\mbox{det}\,\,g=1~$ and 
parametrising the WZNW field by 
\begin{equation}\label{g=c,v_n}
  g=cI+v^n~t_n= \left( \begin{array}{cr}
  c-v_2&-v_1-v_0\\-v_1+v_0&c+v_2 \end{array}\right),~~~~~\mbox{with}~~~~~
c^2+v^nv_n=1,
\end{equation}
for each of the coset models only a couple of these field components
will be gauge invariant.

It is worth to consider both Lagrangean and Hamiltonian reduction.

\subsection{Lagrangean reduction by gauging}

The considered $SL(2,R)$ WZNW theory on the cylinder can be gauged in
the standard manner and it yields for the compact axial respectively
vector cases the gauge invariant  Lagrangeans with the euclidian target
space geometries of a cigar (the `euclidian black hole') and a trumpet 
\begin{equation}\label{Compact}
{\mathcal L}_G^{(1)} |=\frac{1}{\gamma^2}\,\,
 \frac{\partial_zv_1 \partial_{\bar z}v_1 +
\partial_zv_2\partial_{\bar z}v_2}{1+v_1^2 + v_2^2},~~~~~~~~
{\mathcal L}_G^{(2)} |=\frac{1}{\gamma^2}\,\,
 \frac{\partial_zc \partial_{\bar z}c +
\partial_zv_0\partial_{\bar z}v_0}{c^2 + v_0^2-1}.
\end{equation}
Whereas the target space of ${\mathcal L}_G^{(1)} |$ is $R^2$,
for ${\mathcal L}_G^{(2)} |$  the unit
disk $c^2+v_0^2<1$ is missing and this Lagrangean 
is singular at the disk boundary, but both coset theories are 
mutually related \cite{FJW2}.

For the non-compact cases we obtain two equivalent minkowskian
('black hole') actions
\begin{equation}\label{Non-compact}
{\mathcal L}_G^{(3)} |=\frac{1}{\gamma^2}\,\,
 \frac{\partial_zv_1 \partial_{\bar z}v_1 -
\partial_zv_0\partial_{\bar z}v_0}{1+v_1^2 - v_0^2},~~~~~~~~
{\mathcal L}_G^{(4)} |=\frac{1}{\gamma^2}\,\,
 \frac{\partial_zv_2 \partial_{\bar z}v_2 -
\partial_zc\partial_{\bar z}c}{1+ v_2^2-c^2 },
\end{equation}
which are analytically related to (\ref{Compact}). The target space is
$R^2$ and ${\mathcal L}_G^{(3)} |$, e.g., has two singularity lines
$~v_0=\pm\sqrt{1+v_1^2}~$. There are so three different regular
domains in the target space
\begin{equation}\label{regular}
v_0>\sqrt{1+v_1^2},~~~-\sqrt{1+v_1^2}<v_0<\sqrt{1+v_1^2},~~~ 
v_0<-\sqrt{1+v_1^2}, 
\end{equation}
and this coset theory has to be investigated in each of them separately.

Finally, for the nilpotent gaugings only two identical gauged
Lagrangeans arise for the field $~V=g_{12}(z, \bar z)~$ whereas
the other gauge invariant components $v_2$ or $c$ simply disappear 
\begin{equation}\label{Nilpotent}
{\mathcal L}_G^{(5)} |={\mathcal L}_G^{(6)} |=\frac{1}{\gamma^2}\,\,
 \frac{\partial_zV \partial_{\bar z}V}{V^2}.
\end{equation}
$V=0$ is a singularity of the Lagrangian, but for the regular parametrisation
$V=\pm e^{\gamma\phi}$ we get as a result free-field theories only
\begin{equation}\label{Free}
{\mathcal L}_G^{(5)} |={\mathcal L}_G^{(6)} |=
 \partial_z\phi \partial_{\bar z}\phi.
\end{equation}

Note that the Liouville theory does not arise by this standard
gauging.

\subsection{Hamiltonian reduction by constraints}

Hamiltonian reduction is an alternative but more flexible method to
construct and investigate coset theories. Here the constrained
Kac-Moody currents $J_0=0=\bar J_0$, $J_2=0=\bar J_2$ and $J_+=0=\bar
J_+$ provide the cosets (\ref{Compact}), (\ref{Non-compact}) and
(\ref{Nilpotent}) respectively. But both, the axial and the vector
gauged Lagrangeans arise by one and the same constraints 
\cite{FJW, FJW2}.  Although these systems are described by different
components of the WZNW field they live on the same constrained surface, and
are therefore mutually related with each other. It is important to
mention that the two first current constraints are of second class and
the nilpotent gauging is of first class.

Imposing to the $SL(2,R)$ WZNW theory the alternative nilpotent
constraints $J_+=\rho,~\bar J_+=\bar\rho~$ with non-vanishing
constants $\rho$ and $\bar\rho$, and write the only gauge
invariant field component as $g_{12}(z, \bar z)=
\psi(z)\bar\psi(\bar z)+\chi(z)\bar\chi(\bar z)$,
then $\psi(z)=g_{11}(z),~ \chi(z)= g_{12}(z)$ etc.,
satisfy constant Wronskians 
$~\psi(z)\chi'(z)-\psi'(z)\chi(z)=\rho\gamma^2$ etc..
The following identification
\begin{equation}\label{Expo}
e^{-\gamma\varphi(z, \bar z)}=\psi(z)\bar\psi(\bar z)+\chi(z)\bar\chi(\bar z)
\end{equation}
leads us to the Liouville equation with the `cosmological' constant given by
$~\mu=-\rho\bar\rho\gamma^3~$
\begin{equation}\label{Lio}
\partial_{z\bar z}\, \varphi +\mu e^{2\gamma\varphi}=0.
\end{equation}
The equation (\ref{Expo}) obviously also provides the general 
solution of the Liouville equation. 

Hamiltonian reduction is
in fact a method for integrating coset theories \cite{FJW}.

\section{Reduction of Poisson brackets}

Since the nilpotent constraints are of first class, for the gauge
invariant Liouville exponential $e^{-\gamma\varphi}=g_{12}(z, \bar z)$
the reduced non-equal time Poisson bracket can be read off directly
from the relation (\ref{PB-g-g}) without any further calculations. For
the (anti-)chiral fields $\psi(z)=g_{11}(z),\bar\psi(\bar z)=\bar
g_{11}(\bar z),\,\,\chi(z)=g_{12}(z),\,\bar\chi(\bar z)=\bar g_{12}(\bar
z)$ the classical form of the exchange algebra results, which quantum
mechanically become the celebrated Gervais-Neveu quantum group
relations.

It might be worth to note that the gauge invariant 
nilpotent reduction of the Sugawara
energy-momentum tensor immediately generates the 
Liouville form with the standard classical 'improvement' term
included. But in this case there do not exist coset currents.

The situation is different if we reduce the $SL(2,R)$ WZNW theory by
the second class constraints. Using the isomorphism between $SL(2,R)$
and $SU(1, 1)$ there is a natural complex structure given, e.g. for
the euclidian case (\ref{Compact}), by the complex coordinates
$u=v_1+iv_2$ and $x=c+iv_0$ which are related by $|x|^2-|u|^2=1$.
$u(z,\bar z)$ and $x(z,\bar z)$ are the physical fields of the axial
respectively vector gauged cosets. These gauge invariant fields can be
expressed similarly as in the Liouville case (\ref{Expo}) by
$~u(z,\bar z)=\psi (z)\bar\psi (\bar z)+\chi (z)\bar\chi (\bar z)~$,
but now in terms of the complex fields
$~\psi(z)=g_{11}(z)+ig_{12}(z)$, $~\chi(z)=g_{21}(z)+ig_{22}(z)~$
etc., and with the non-constant Wronskians
$~\psi(z)\chi'(z)-\psi'(z)\chi(z)=2W(z)~$ and the anti-chiral one.
Here $~W(z)=J_1(z)+iJ_2(z)~$ is the parafermionic coset current
\cite{MW, FJW}.  The algebra of the coset fields is given by
Dirac brackets \cite{FJW2}, and there are causal relations for each
 coset
\begin{eqnarray}\label{DB-u-u}
\{\,u(z,\bar z),\, u(y,\bar y)\,\}_D&=&\gamma^2\,
\,\Theta\,
[u(z,\bar y)\,u(y,\bar z)-
u(z,\bar z)\,u(y,\bar y)],\nonumber \\
\{\,u(z,\bar z),\, u^*(y,\bar y)\,\}_D&=&\gamma^2\,
\,\Theta\,
 x(z,\bar y)\,x^*(y,\bar z),\nonumber\\
\{\,x(z,\bar z),\, x(y,\bar y)\,\}_D&=&\gamma^2\,
\,\Theta\,
[x(z,\bar y)\,x(y,\bar z)-
x(z,\bar z)\,x(y,\bar y)],\nonumber\\
\{\,x(z,\bar z),\, x^*(y,\bar y)\,\}_D&=&\gamma^2\,
\,\Theta\,
u(z,\bar y)\,u^*(y,\bar z),
\end{eqnarray}
and non-causal connections which with the  notation
$2E =\epsilon(z-y)$,  $2\bar E =\epsilon(\bar z-\bar y)$
are
\begin{eqnarray}
\label{DB-u-x}
\{\,u(z,\bar z),\, x(y,\bar y)\,\}_D&=&\gamma^2\,
\,\Theta\,
x(z,\bar y)\,u(y,\bar z)-
\gamma^2\, E\,u(z,\bar z)\,x(y,\bar y),\nonumber\\
\{\,u(z,\bar z),\, x^*(y,\bar y)\,\}_D&=&\gamma^2\,
\Theta\,
u(z,\bar y)\,x^*(y,\bar z)-
\gamma^2\,\bar E\, u(z,\bar z)\,x^*(y,\bar y).
\end{eqnarray}

As expected the axial and vector gauged theories form a
coupled algebra.

\section{Canonical quantisation}

The canonical quantisation of the cosets can be performed in the same
way as it has been done for the Liouville theory \cite{OW}. Here one
uses the general solution of the coset as a canonical transformation
between the non-linear coset fields and free fields. The quantisation
will be defined by replacing the Poisson brackets of the canonical
free fields by commutators. Non-linear expressions in the free fields
will be normal ordered. But calculations with normal ordered operators
usually yield anomalous contributions. Such anomalies can be avoided
by quantum mechanically deforming the composite operators of the
cosets. The deformations are determined by requiring the classical
symmetry transformations, and locality, to be valid as commutator
relations. As a result, we show the non-equal time commutator for the
Liouville exponential $u(z, \bar z)=e^{-\gamma\varphi(z, \bar z)}$
which is written here for convenience as a Moyal bracket \cite{JW}
\begin{eqnarray}\label{u*u!}
\{\check u(z,\bar z),\,\check u(y,\bar y)\}_*=
\frac{1}{\hbar}\,\sin(\hbar\gamma^2/4)\,
[\epsilon(z-y)+\epsilon(\bar z-\bar y)]\times
~~~~~~~~~~~~~~~~~~~~~
\\ \nonumber
\,\left[\check u(z, \bar y) * \check u(y, \bar z) +\check u(y, \bar z)*
\check u(z, \bar y)-
\frac{\check u(z, \bar z)* \check u(y, \bar y)+\check u(y, \bar y)*
\check u(z, \bar z)}{2\cos(\hbar\gamma^2/4)}
\right].
\end{eqnarray}
Its expansion in $\hbar$ reproduces the 
Poisson bracket. We can define by (\ref{u*u!}) other operators and 
their commutators. 
The operator of $e^{-2\gamma\varphi(z,\bar z)}$ simply follows
by differentiation for equal time.

Further results will be discussed in the lecture notes \cite{JW2}.

\section{Final remarks}

There is a complete classical understanding of the whole set of
$SL(2,R)$ theories. Quantum mechanically the Liouville theory is best
worked out, but still incomplete. The zero mode structure and the
Hilbert space require further intensive study. Coset currents only exist for
the non-nilpotent gauged $SL(2,R)$ theories as parafermions. They generate
quantum mechanically a dilaton which might render the classically
non-dynamical metric dynamical.

Besides being interesting in its own right the $SL(2,R)$ WZNW model
and its cosets also appear in many applications, in particular in
string calculations. An exact and complete quantum mechanically
treatment of the $SL(2,R)$ family would be helpful to understand,
e.g., the $AdS_3/CFT$ correspondence, which is an intensively
discussed  contemporary problem.

Minor  knowledge exists for the quantum mechanical
$SL(2,R)$ WZNW model.

\end{document}